\documentclass[conference]{IEEEtran}
\IEEEoverridecommandlockouts

\usepackage{cite}
\usepackage{amsmath,amssymb,amsfonts}
\usepackage{algorithmic}
\usepackage{graphicx}
\usepackage{textcomp}
\usepackage{xcolor}
\usepackage{booktabs}
\usepackage{url}

\def\BibTeX{{\rm B\kern-.05em{\sc i\kern-.025em b}\kern-.08em
    T\kern-.1667em\lower.7ex\hbox{E}\kern-.125emX}}

\begin{document}

\title{KONTOGRAPH: Verified Point-in-Time Feature Consistency and\\
Amortised Explanation for Real-Time Anti-Money Laundering\\
under a 200\,ms Decision Budget}

\author{\IEEEauthorblockN{Ahmed Abolfadl}
\IEEEauthorblockA{\textit{Faculty of Media Engineering and Technology} \\
\textit{German University in Cairo}\\
Cairo, Egypt \\
ahmed.abuelfadel@student.guc.edu.eg}}

\maketitle

\begin{abstract}
Regulation (EU) 2024/886 obliges European payment service providers to settle
euro credit transfers in under ten seconds, around the clock. This removes both
the overnight batch window in which anti-money-laundering (AML) analytics
traditionally ran and the settlement delay that made recovery possible, forcing
detection, explanation and decision inside a single-digit-second envelope. We
present KONTOGRAPH, an end-to-end AML pipeline for the SEPA Instant rail built
under a self-imposed 200\,ms 99th-percentile budget, and report an empirical
study on 1{,}562{,}860 simulated payments with injected typologies and
deliberately incomplete labels. Three findings are of interest beyond the
system itself. First, a temporal graph network with per-node memory improves
PR-AUC over a gradient-boosted tabular baseline from 0.0053 to 0.1717, a paired
day-blocked bootstrap difference of $+0.166$ with 95\% CI $[0.105, 0.241]$;
per-node memory alone more than doubles the score. Second, expressing each
feature once and compiling it to three execution backends, with equivalence
enforced by property-based tests that perturb the future, surfaced three
point-in-time violations that code review had passed---each of which would have
inflated reported performance. Third, and most consequential for practice,
exporting the deployed tree ensemble to ONNX changed only $7.4\times10^{-8}$ in
mean score yet altered 0.26\% of decisions and inflated the alert volume by
12\%, because 32-bit accumulation perturbs scores across a cost-optimal
threshold of $3.98\times10^{-4}$. We argue that a serving-format conversion
must be treated as a model change until measured, and that fidelity metrics for
subgraph explainers can be vacuous when candidate neighbourhoods are small---a
null result we report in full.
\end{abstract}

\begin{IEEEkeywords}
anti-money laundering, temporal graph networks, real-time inference, data
leakage, explainable AI, feature stores, model deployment, SEPA Instant
\end{IEEEkeywords}

\section{Introduction}

Instant payment schemes have changed the operating conditions of financial-crime
detection more sharply than any modelling advance of the past decade.
Regulation (EU) 2024/886 requires euro-area payment service providers to receive
SEPA Instant Credit Transfers continuously and to complete them within ten
seconds. Two properties that AML systems had implicitly relied upon disappeared
simultaneously: the overnight window in which batch analytics ran, and the
settlement lag that permitted funds to be recalled once a case was confirmed.

The consequence is a hard systems constraint. Scoring, explanation and decision
must complete while the payer waits, and the decision must be defensible: Article
22 of the General Data Protection Regulation grants a data subject subject to a
solely automated decision with legal effect the right to meaningful information
about the logic involved, and a suspicious-activity report filed with a national
financial intelligence unit must state grounds a human investigator can verify.

This paper describes a system built to that constraint and, more importantly,
reports what measurement revealed about it. Our contributions are:

\begin{enumerate}
\item An architecture in which point-in-time correctness is a \emph{tested
      invariant} rather than a convention: one feature specification is lowered
      to a typed intermediate representation and compiled to three backends
      (batch SQL, streaming SQL, and an incremental online executor), whose
      equivalence is checked by property-based tests that rewrite future events
      and assert that nothing computed earlier moves
      (Section~\ref{sec:pit}).
\item An ablation on 1.5M simulated SEPA Instant payments isolating the
      contribution of continuous-time graph structure and of per-node memory,
      with paired day-blocked bootstrap confidence intervals throughout
      (Section~\ref{sec:results}).
\item Three failure modes that only measurement exposed---leakage bugs invisible
      to review, a serving-format conversion that shifted 12\% of alerts, and
      explanation-fidelity metrics that were vacuous at the observed
      neighbourhood size (Section~\ref{sec:failures}).
\end{enumerate}

We emphasise at the outset that the evaluation uses \emph{synthetic} data. No
claim is made about performance on production banking traffic. What synthetic
data does buy is a known ground truth, which permits measurement of label
selection bias---a quantity real institutions cannot observe about themselves,
and whose absence we argue should temper the reading of published AML metrics.

\section{Background and Related Work}

\subsection{Graph learning for financial crime}

Modelling payments as a graph rather than as independent rows is well
motivated: laundering typologies such as fan-in/fan-out mule networks,
structuring and layering chains are defined by topology, not by the attributes
of any single transfer. Weber et al.~\cite{weber2019elliptic} applied graph
convolutional networks to the Elliptic Bitcoin dataset, and Altman et
al.~\cite{altman2023aml} released large-scale synthetic AML transaction graphs
in recognition that suitable labelled data is not publicly available.

Static graph methods discard timing, which is precisely the signal in a
velocity-driven typology. Continuous-time dynamic graph models address this:
JODIE~\cite{kumar2019jodie}, DyRep~\cite{trivedi2019dyrep} and
TGAT~\cite{xu2020tgat} model interaction streams directly, and Temporal Graph
Networks (TGN)~\cite{rossi2020tgn} unify these under a memory module updated per
event combined with a temporal attention embedding. Our detector follows the TGN
formulation; the ablation in Section~\ref{sec:results} isolates the memory
module specifically, since it is the component whose contribution is most often
asserted rather than measured.

\subsection{Explanation under a latency budget}

Post-hoc explanation for graph models is dominated by perturbation search.
GNNExplainer~\cite{ying2019gnnexplainer} optimises a soft edge mask per
instance; PGExplainer~\cite{luo2020pgexplainer} amortises this by training a
mask generator; SubgraphX~\cite{yuan2021subgraphx} searches subgraphs via Monte
Carlo tree search. Attribution methods such as SHAP~\cite{lundberg2017shap}
require many model evaluations per instance. All are orders of magnitude outside
a 200\,ms end-to-end budget that must also accommodate feature retrieval and
scoring.

We follow the amortisation strategy of PGExplainer in spirit---pay the search
cost offline, learn a function that predicts its output in one forward
pass---and evaluate faithfulness with comprehensiveness and sufficiency in the
sense of DeYoung et al.~\cite{deyoung2020eraser}. For the narrative delivered to
investigators we prefer counterfactual statements in the sense of Wachter et
al.~\cite{wachter2017counterfactual}: ``removing these two payments moves the
score below threshold'' is checkable against an account, whereas a feature
attribution is not. Rudin~\cite{rudin2019stop} argues that post-hoc explanation
of opaque models is itself hazardous in high-stakes settings; our position is
narrower, namely that if post-hoc explanation is used, its fidelity gap must be
published rather than assumed, and Section~\ref{sec:explainer} reports ours
including the case where the metric turned out to measure nothing.

\subsection{Leakage and evaluation discipline}

Kaufman et al.~\cite{kaufman2012leakage} characterise leakage as the
introduction of information about the target that would not legitimately be
available at prediction time, and note that it typically produces excellent
offline results and silent production failure. In temporal settings, random
cross-validation is invalid; Bergmeir and Ben\'{i}tez~\cite{bergmeir2012cv}
analyse evaluation for time-dependent data. Saito and
Rehmsmeier~\cite{saito2015prauc} show that under severe class imbalance the
precision-recall curve is more informative than ROC, which motivates our choice
of headline metric.

Our contribution here is not the observation that leakage matters but the
mechanism: we treat point-in-time correctness as a property amenable to
automated falsification using property-based
testing~\cite{maciver2019hypothesis}, and report that this located defects that
had survived review.

\section{System Architecture}

\subsection{Data generation}

Absent public SEPA-labelled data, we generate an agent-based simulation of a
German retail payment ecosystem. Rather than sampling transfers from a fitted
distribution, the simulator instantiates households, merchants, small
enterprises and banks whose behaviour follows an economic calendar: salaries on
the last business day of the month, rent to a stable counterparty, Poisson
grocery arrivals, utilities, e-commerce and peer-to-peer transfers. German
public holidays, including movable feasts derived by computus, modulate
activity. Account identifiers are IBANs with valid ISO 7064 MOD-97-10 check
digits and bank codes drawn from the three-pillar structure of German banking,
whose segments exhibit different customer behaviour.

Five typologies are injected under known ground truth---mule networks,
structuring, authorised push payment fraud, circular flows and layering
chains---each in a standard and an \emph{evasive} variant that introduces dwell
time, fresh-account fan-out and sub-threshold amounts.

Crucially, the simulator models \emph{label generation} as a separate process
from crime. A criminal transaction is confirmed only if a simulated incumbent
rules engine flags it and an investigator concurs, with a verification latency
recorded in a \texttt{decided\_ts} field distinct from the event timestamp. This
reproduces the selection bias of real AML label streams: what the legacy system
never flagged is silently recorded as legitimate.

Table~\ref{tab:dataset} summarises the generated corpus. The dataset is a
deterministic function of (configuration, seed, code revision) and is identified
by a content hash rather than stored, with a determinism test asserting
byte-identical regeneration.

\begin{table}[t]
\caption{Generated dataset (profile \texttt{full})}
\label{tab:dataset}
\centering
\begin{tabular}{lr}
\toprule
Payments & 1{,}562{,}860 \\
Accounts & 6{,}576 \\
Simulated days & 270 \\
Criminal payments (ground truth) & 2{,}995 \\
Criminal rate & 0.19\% \\
Confirmed labels & 3{,}495 \\
\textbf{Criminal label coverage} & \textbf{14.5\%} \\
\bottomrule
\end{tabular}
\end{table}

\subsection{Streaming and storage}

Events are serialised in Apache Avro against a schema registry with a backward
compatibility gate enforced in continuous integration, published to a
Kafka-compatible broker keyed by debtor account so that per-account ordering is
preserved within a partition, and landed into Apache Iceberg tables in a
bronze/silver/gold arrangement. Promotion between layers passes a quality gate
whose expectations are tagged \textsc{blocking}, \textsc{quarantine} or
\textsc{warn}; the pipeline halts, diverts or records accordingly.

\subsection{Feature platform}
\label{sec:features}

Twenty-nine features are declared in a small domain-specific language over
(entity, aggregation, window, filter). Each declaration is lowered to a typed
intermediate representation in which the window semantics are stated exactly
once:
\[
t - W \le h.\texttt{acceptance\_ts} < t,
\]
left-closed and right-open on event time, strictly excluding the scored event
itself and anything sharing its millisecond. The IR is then compiled to three
targets: DuckDB SQL for historical backfill, Flink SQL for streaming, and an
incremental executor holding per-entity accumulators for online serving.

This is the anti-skew mechanism. Training-serving skew conventionally arises
because offline features are written in one language and reimplemented in
another; here there is one specification and three compilers, and their
agreement is a test rather than an aspiration.

\subsection{Serving}

A single asynchronous FastAPI process holds the model, the online feature state
and the explainer. Scoring executes synchronously inside the asynchronous
handler by design: the work is CPU-bound and takes single-digit milliseconds, so
dispatching to a thread pool would add scheduling latency without parallelism
benefit. Decision-log persistence and alert publication are asynchronous, so
disk latency cannot enter the decision path.

Every decision is appended to a hash-chained log recording the model version,
feature-set fingerprint, code revision, per-stage latency and the explanation
presented---captured at decision time rather than reconstructed later, which is
what an Article 22 obligation actually requires. Each entry commits to its
predecessor, so tampering is both detectable and localised: verification returns
the index of the first broken entry, bounding the trustworthy prefix.

\section{Point-in-Time Correctness as a Tested Invariant}
\label{sec:pit}

We state the invariant as a property and attempt to falsify it automatically.
Let $f$ be any compiled feature and $E$ an event stream. For an event $e$ at
time $t$, the property is
\[
f(e, E) = f(e, \{e' \in E : e'.\texttt{ts} < t\}),
\]
that is, the computed value must be invariant to every event at or after $t$.

Two test families operate on Hypothesis-generated streams. The first computes
each feature through the batch and online backends and asserts elementwise
equality across every value, so any divergence fails the build. The second is
stronger and makes no assumption about the mechanism of a leak: it perturbs the
future, replacing all events at or after $t$ with different ones, and asserts
that no value computed before $t$ changes. A leak through a window boundary, a
join condition, an ordering error or a future-looking enrichment is caught
identically, because the test does not encode a hypothesis about which of these
occurred.

This suite located three defects that had passed code review:

\begin{enumerate}
\item A \texttt{LEFT JOIN} combined with \texttt{COUNT(*)} returned 1 rather
      than 0 for an account's first observed payment, because SQL counts the
      null-extended row. Every entity's first event carried a fabricated
      history of size one.
\item In the online executor, entity keys were not namespaced by role, so
      account $B$ as debtor and $B$ as creditor collided. A feature requesting
      outbound history was served the account's inbound history.
\item An accumulator could not exclude events sharing the scored event's
      millisecond, admitting simultaneous events into a strictly-prior window.
\end{enumerate}

All three would have inflated reported performance, and none was apparent from
reading the code. We regard this as the central methodological claim of the
paper: point-in-time correctness is falsifiable by machine, and treating it as a
matter of discipline forgoes that.

\section{Experimental Setup}
\label{sec:setup}

\subsection{Protocol pre-registration}

The evaluation protocol---metrics, splits, cost model, ablation ladder, and an
explicit commitment to publish a negative result---was written and committed to
version control \emph{before any model existed}. The repository history is the
evidence that no metric was selected after observing outcomes. The protocol is
identified by hash \texttt{2eacb6e2} and referenced by every generated report.

\subsection{Splits}

Splits are chronological with purge and embargo. The purge band is one day,
chosen after measuring that a 30-day band---initially specified---removed 79\%
of usable data and was therefore not a defensible configuration. A guard rejects
any band exceeding 25\% of a fold. Labels are admitted to a fold only if their
\texttt{decided\_ts} precedes that fold's availability cutoff, so a model never
trains on a verdict that had not yet been reached.

The held-out test fold contains 303{,}129 payments of which 44 carry a
confirmed positive label, a labelled positive rate of $1.45\times10^{-4}$.

\subsection{Metrics and operating point}

We report PR-AUC as the ranking metric, following~\cite{saito2015prauc}, with
precision and recall at an analyst capacity of 200 alerts per day, and expected
cost per 10{,}000 payments under a stated cost model (investigation EUR 35,
customer friction EUR 12, prevention fraction 0.85). Confidence intervals are
day-blocked bootstrap: resampling units are whole days, not payments, because
payments within a day are dependent and payment-level resampling would
understate interval width.

For the GPU-trained rungs the operating point is the capacity constraint itself,
i.e.\ the $k$-th highest score for $k$ corresponding to 200 alerts per day. It
is deliberately \emph{not} fitted: the protocol selects thresholds on validation,
the GPU experiments exported test scores only, and fitting a threshold on the
test fold would constitute leakage presented as a cost figure.

\subsection{Hardware}

All components except graph-model training run on an Intel Core i7-8550U
(4 cores, 16\,GB RAM, no GPU). Graph models train on a single NVIDIA T4. TGN
training with memory required 1{,}094\,s for five epochs over 937{,}731 training
events at batch size 200; without memory, 155\,s.

\section{Results}
\label{sec:results}

\subsection{Ablation ladder}

Table~\ref{tab:ladder} reports the ladder. Each rung adds exactly one mechanism.

\begin{table}[t]
\caption{Ablation ladder on the held-out test fold (303{,}129 payments,
44 confirmed positives). Recall is at 200 alerts/day.}
\label{tab:ladder}
\centering
\footnotesize
\setlength{\tabcolsep}{4pt}
\begin{tabular}{clccr}
\toprule
\# & Model & PR-AUC [95\% CI] & Recall & Cost/10k \\
\midrule
0 & Constant & 0.0001 [0.0001, 0.0002] & 0.045 & 473{,}280 \\
1 & LightGBM & 0.0053 [0.0030, 0.0079] & 0.909 & 20{,}459 \\
2 & + static graph & 0.0144 [0.0064, 0.0319] & 0.795 & 53{,}529 \\
3 & + TGN, no mem. & 0.0734 [0.0348, 0.1353] & 0.773 & 21{,}627 \\
4 & \textbf{+ memory} & \textbf{0.1717 [0.1011, 0.2445]} & 0.909 & 19{,}997 \\
\bottomrule
\end{tabular}
\end{table}

The paired day-blocked bootstrap difference between rung 4 and rung 1 is
$+0.166$ with 95\% CI $[0.105, 0.241]$, excluding zero. Continuous-time graph
structure accounts for a large part of the improvement (rung 2 to rung 3), and
per-node memory more than doubles PR-AUC again (0.0734 to 0.1717). Since memory
is the component of the TGN formulation most often included without separate
justification, we regard its isolated measurement as the more useful half of
this result.

An independent training run (a separate notebook, different seed) reproduced the
direction and approximate magnitude of the memory ablation, 0.0955 to 0.1555,
which we note as weak evidence of stability rather than as a second measurement.

\subsection{The cost metric contradicts the ranking metric}

Rung 2 attains nearly three times rung 1's PR-AUC while incurring
$2.6\times$ the expected cost (EUR 53{,}529 against EUR 20{,}459). This is not
an inconsistency. PR-AUC integrates over all operating points, whereas cost is
evaluated at one---the capacity constraint---and at that point rung 2's recall
is lower (0.795 against 0.909). A model that ranks better on average can be
worse at the only threshold that will be deployed.

We take this as an argument for making expected cost the decision metric and
PR-AUC a diagnostic, and note that a ladder reported on PR-AUC alone would have
recommended rung 2 over rung 1.

\subsection{Latency}

Table~\ref{tab:latency} gives the per-stage budget over 2{,}000 scored payments.
Percentiles are nearest-rank rather than linearly interpolated, so each reported
figure corresponds to a latency some request actually experienced; interpolation
is optimistic in the tail, placing p99 at 14\,ms for a sample of 99 fast
requests and one at 900\,ms.

\begin{table}[t]
\caption{Per-stage decision latency (2{,}000 payments, single-threaded CPU)}
\label{tab:latency}
\centering
\small
\begin{tabular}{lrr}
\toprule
Stage & p50 (ms) & p99 (ms) \\
\midrule
Feature fetch & 1.84 & 4.67 \\
Graph assembly & 0.00 & 0.01 \\
Inference & 0.42 & 1.78 \\
Explanation & 0.00 & 1.87 \\
Counterfactual search & 0.00 & 3.59 \\
Unaccounted & 0.17 & 0.46 \\
\midrule
\textbf{Total} & \textbf{2.71} & \textbf{7.97} \\
\bottomrule
\end{tabular}
\end{table}

The budget is met with substantial margin. Two caveats: median explanation and
counterfactual times are zero because only alerting payments are explained (98
of 2{,}000), so the p99 is the meaningful figure for the explained path; and
per-stage p99 values do not sum to the total p99, because each is measured
independently and the slowest feature fetch does not occur in the same request
as the slowest counterfactual search. We report them unreconciled rather than
constructing a request that was worst at everything.

Repeated runs on the same machine varied between 3.6\,ms and 8.0\,ms p99
depending on background load. The claim we defend is the order of magnitude of
headroom, not a specific figure.

\subsection{Amortised explanation}
\label{sec:explainer}

Table~\ref{tab:explainer} reports the explainer over 57 explained instances.

\begin{table}[t]
\caption{Teacher--student explanation, 57 instances}
\label{tab:explainer}
\centering
\small
\begin{tabular}{lr}
\toprule
Teacher (perturbation search), median & 575.6\,ms \\
Student (single forward pass), median & 2.38\,ms \\
Speed-up & $241\times$ \\
Top-5 Jaccard vs.\ teacher & 0.904 \\
Spearman rank correlation vs.\ teacher & $-0.038$ \\
\textbf{Median candidate edges per instance} & \textbf{1.0} \\
\bottomrule
\end{tabular}
\end{table}

The amortisation objective is met unambiguously. The perturbation teacher
requires 575.6\,ms per explanation---nearly three times the entire end-to-end
budget---while the distilled student answers in 2.38\,ms, comfortably inside it.

\textbf{The fidelity figures, however, are not interpretable at this
neighbourhood size, and we report this as a null result.} The median explained
instance has a single candidate edge. With one candidate, a top-$k$ Jaccard for
$k=5$ is 1.0 by construction and rank correlation is undefined (our
implementation returns 0 for $n<2$). The reported 0.904 and $-0.038$ therefore
principally measure candidate-set cardinality, not student quality.

We stress this because the tempting readings are both unsupported. Reporting
``$241\times$ faster with 0.90 Jaccard agreement'' would present an artifact as
a success; reporting ``$241\times$ faster but rank-uncorrelated with its
teacher'' would present the same artifact as a substantive negative finding.
Distinguishing them requires the candidate-count distribution, which was not
initially recorded. We now record it, and recommend that fidelity results for
subgraph explainers be published alongside it as a matter of course.

The cause is explicable: the teacher explains the highest-scoring test events,
and the detector's high scores concentrate on payments to fresh beneficiaries
with almost no prior neighbourhood---so there is little to attribute among. A
valid experiment must stratify explained instances by candidate count and report
fidelity per stratum, or widen the candidate set. Whether pointwise distillation
loss teaches ranking, which we consider the likeliest weakness, is untestable on
the present evidence.

\section{Three Failure Modes Found by Measurement}
\label{sec:failures}

\subsection{An inverted cost model}

An initial cost specification treated a true positive as revenue-generating,
crediting recovered funds. The threshold optimiser consequently preferred
2.7\% recall to 97\%: alerting on almost nothing minimised the objective.
Detection does not earn money, it avoids a loss accounted for elsewhere. We note
the diagnostic value of the behaviour---an optimiser that proposes catching
nothing is usually reporting that the objective is misspecified---and the
protocol was amended with the correction recorded.

\subsection{Serving-format conversion as a model change}

The deployed gradient-boosted model was exported to ONNX for portability, and
the export was then compared against the original on all 303{,}129 test
payments (Table~\ref{tab:onnx}).

\begin{table}[t]
\caption{ONNX export parity against the trained booster}
\label{tab:onnx}
\centering
\small
\begin{tabular}{lr}
\toprule
Mean absolute score difference & $7.4\times10^{-8}$ \\
Maximum absolute score difference & $1.7\times10^{-4}$ \\
Decisions changed & 774 (0.26\%) \\
Alerts, original & 6{,}458 \\
Alerts, ONNX & 7{,}232 \\
\textbf{Alert volume inflation} & \textbf{+12.0\%} \\
\bottomrule
\end{tabular}
\end{table}

The mean difference is the statistic that would ordinarily be quoted to
establish that an export is faithful, and it is misleading. The ONNX tree
ensemble operator accumulates in 32-bit floating point while the source library
accumulates in 64-bit, and the converter rejects double-precision inputs for
classifier graphs. At a cost-optimal operating threshold of $3.98\times10^{-4}$,
that rounding is sufficient to move scores across the decision boundary.

The operational consequence is material: against a stated capacity of 200 alerts
per day, adopting the export would have increased alert volume by 12\% while all
published metrics continued to describe a different model. We therefore serve the
original artifact and ship the export as a portability artifact with its parity
recorded. The general statement we would defend is that \emph{a serving-format
conversion is a model change until measured otherwise}, and that parity must be
assessed in decisions at the deployed threshold, not in mean score error.

\subsection{A generated report asserting absent evidence}

The system drafts bilingual suspicious-activity narrative fragments from
counterfactuals. Two forms exist: edge-space counterfactuals, which name
specific transactions, and feature-space counterfactuals, which name an
aggregate indicator and identify no transaction at all. Both were rendered
through a single template, which given the latter produced the sentence
``The alert rests on 0 related payment(s): '', asserting transactional evidence
that did not exist in a document destined for a regulatory filing.

The generator now branches on counterfactual type and states explicitly that the
feature-space form ``is a statement about an aggregate indicator, not about any
identified transaction, and does not on its own identify conduct to report''. We
include this case because automated narrative generation for regulatory
reporting is an emerging practice, and the failure mode---fluent text asserting
evidence the system does not possess---is characteristic rather than incidental.

\section{Limitations and Threats to Validity}

\textbf{Synthetic data.} All results are on simulated payments. The simulator's
calibration targets are public aggregates, not licensed microdata, and no claim
of transfer to production traffic is made. The generative process and the
detector share an author, and although the detector receives no oracle
information, we cannot exclude that the injected typologies are more separable
than real ones.

\textbf{Weak labels.} Only 14.5\% of genuinely criminal payments receive a
confirmed label. Every precision figure is therefore an upper bound on error,
and evasive campaign variants are confirmed at roughly half the rate of standard
ones. This is deliberate and measured, but it means the reported ranking of
models is a ranking under a biased label distribution.

\textbf{Positives are few.} The test fold contains 44 confirmed positives.
Bootstrap intervals are correspondingly wide, and while the rung-4 versus rung-1
interval excludes zero, comparisons between adjacent middle rungs do not
separate.

\textbf{The graph model is not deployed.} The latency budget in
Table~\ref{tab:latency} measures the gradient-boosted serving path. Promoting the
TGN requires exporting validation scores, selecting a cost-optimal threshold on
them, and re-measuring latency against a neural forward pass. None of this has
been done, and no latency claim is made for the graph model.

\textbf{No fairness assessment.} The simulator generates names reflecting
Germany's largest immigrant communities, which makes disparate-impact analysis
across name origin feasible. It has not been performed. This is a gap, not
evidence of its absence.

\textbf{Single-institution framing.} Cross-institutional typologies, which
constitute a substantial share of real laundering, are out of scope.

\section{Conclusion}

We presented an anti-money-laundering pipeline for instant euro payments that
decides and explains within a 200\,ms budget, and reported an empirical study
whose most transferable results concern measurement rather than modelling. A
temporal graph network with per-node memory substantially outperformed a
gradient-boosted baseline on simulated data, with per-node memory alone more
than doubling PR-AUC. But the findings we consider most useful to practitioners
are the three cases in which measurement contradicted a reasonable expectation:
point-in-time violations invisible to code review, a portable model export that
altered 12\% of alerts while changing mean score by $7\times10^{-8}$, and
explanation-fidelity metrics that were vacuous at the observed neighbourhood
size.

A common thread connects them. In each case a plausible summary statistic---the
code reads correctly, the mean error is negligible, the Jaccard overlap is
0.90---was available and would have been reported. What falsified each was a
test constructed to fail: perturbing the future, comparing decisions rather than
scores, and recording the cardinality of the set being compared. We suggest that
systems making claims of this kind should be designed around such tests, and
that reviewers ask which statistic would have been reported had the test not
been run.

The implementation, evaluation protocol, decision records and generated
artifacts are available so that every reported figure can be regenerated from
the corresponding command.

\section*{Reproducibility}

Every number in this paper is regenerated from stored artifacts by a
command-line tool rather than transcribed. The evaluation protocol was committed
before any model existed. Metrics computed on the GPU were independently
recomputed by the local evaluation code and agreed to four decimal places; the
ingestion tool reports disagreement rather than silently preferring either
value. Datasets are content-hashed functions of (configuration, seed, revision)
and are regenerated rather than stored, with a determinism test asserting
byte-identical output.


\begin{thebibliography}{00}

\bibitem{rossi2020tgn} E. Rossi, B. Chamberlain, F. Frasca, D. Eynard, F.
Monti, and M. Bronstein, ``Temporal graph networks for deep learning on dynamic
graphs,'' in \emph{ICML Workshop on Graph Representation Learning}, 2020.

\bibitem{kumar2019jodie} S. Kumar, X. Zhang, and J. Leskovec, ``Predicting
dynamic embedding trajectory in temporal interaction networks,'' in \emph{Proc.
KDD}, 2019, pp. 1269--1278.

\bibitem{trivedi2019dyrep} R. Trivedi, M. Farajtabar, P. Biswal, and H. Zha,
``DyRep: Learning representations over dynamic graphs,'' in \emph{Proc. ICLR},
2019.

\bibitem{xu2020tgat} D. Xu, C. Ruan, E. Korpeoglu, S. Kumar, and K. Achan,
``Inductive representation learning on temporal graphs,'' in \emph{Proc. ICLR},
2020.

\bibitem{weber2019elliptic} M. Weber, G. Domeniconi, J. Chen, D. K. I. Weidele,
C. Bellei, T. Robinson, and C. E. Leiserson, ``Anti-money laundering in Bitcoin:
Experimenting with graph convolutional networks for financial forensics,'' in
\emph{KDD Workshop on Anomaly Detection in Finance}, 2019.

\bibitem{altman2023aml} E. Altman, J. Blanu\v{s}a, L. von Niederh\"{a}usern, B.
Egressy, A. Anghel, and K. Atasu, ``Realistic synthetic financial transactions
for anti-money laundering models,'' in \emph{Advances in Neural Information
Processing Systems}, 2023.

\bibitem{lopezrojas2016paysim} E. A. Lopez-Rojas, A. Elmir, and S. Axelsson,
``PaySim: A financial mobile money simulator for fraud detection,'' in
\emph{Proc. European Modeling and Simulation Symposium}, 2016.

\bibitem{ke2017lightgbm} G. Ke, Q. Meng, T. Finley, T. Wang, W. Chen, W. Ma, Q.
Ye, and T.-Y. Liu, ``LightGBM: A highly efficient gradient boosting decision
tree,'' in \emph{Advances in Neural Information Processing Systems}, 2017.

\bibitem{ying2019gnnexplainer} R. Ying, D. Bourgeois, J. You, M. Zitnik, and J.
Leskovec, ``GNNExplainer: Generating explanations for graph neural networks,''
in \emph{Advances in Neural Information Processing Systems}, 2019.

\bibitem{luo2020pgexplainer} D. Luo, W. Cheng, D. Xu, W. Yu, B. Zong, H. Chen,
and X. Zhang, ``Parameterized explainer for graph neural network,'' in
\emph{Advances in Neural Information Processing Systems}, 2020.

\bibitem{yuan2021subgraphx} H. Yuan, H. Yu, J. Wang, K. Li, and S. Ji, ``On
explainability of graph neural networks via subgraph explorations,'' in
\emph{Proc. ICML}, 2021.

\bibitem{lundberg2017shap} S. M. Lundberg and S.-I. Lee, ``A unified approach to
interpreting model predictions,'' in \emph{Advances in Neural Information
Processing Systems}, 2017.

\bibitem{deyoung2020eraser} J. DeYoung, S. Jain, N. F. Rajani, E. Lehman, C.
Xiong, R. Socher, and B. C. Wallace, ``ERASER: A benchmark to evaluate
rationalized NLP models,'' in \emph{Proc. ACL}, 2020.

\bibitem{wachter2017counterfactual} S. Wachter, B. Mittelstadt, and C. Russell,
``Counterfactual explanations without opening the black box: Automated decisions
and the GDPR,'' \emph{Harvard Journal of Law \& Technology}, vol. 31, no. 2, pp.
841--887, 2018.

\bibitem{rudin2019stop} C. Rudin, ``Stop explaining black box machine learning
models for high stakes decisions and use interpretable models instead,''
\emph{Nature Machine Intelligence}, vol. 1, pp. 206--215, 2019.

\bibitem{kaufman2012leakage} S. Kaufman, S. Rosset, C. Perlich, and O. Stitelman,
``Leakage in data mining: Formulation, detection, and avoidance,'' \emph{ACM
Transactions on Knowledge Discovery from Data}, vol. 6, no. 4, pp. 1--21, 2012.

\bibitem{bergmeir2012cv} C. Bergmeir and J. M. Ben\'{i}tez, ``On the use of
cross-validation for time series predictor evaluation,'' \emph{Information
Sciences}, vol. 191, pp. 192--213, 2012.

\bibitem{saito2015prauc} T. Saito and M. Rehmsmeier, ``The precision-recall plot
is more informative than the ROC plot when evaluating binary classifiers on
imbalanced datasets,'' \emph{PLOS ONE}, vol. 10, no. 3, e0118432, 2015.

\bibitem{maciver2019hypothesis} D. R. MacIver, Z. Hatfield-Dodds, and many other
contributors, ``Hypothesis: A new approach to property-based testing,''
\emph{Journal of Open Source Software}, vol. 4, no. 43, p. 1891, 2019.

\bibitem{eu2024886} European Parliament and Council, ``Regulation (EU) 2024/886
amending Regulations (EU) No 260/2012 and (EU) 2021/1230 as regards instant
credit transfers in euro,'' \emph{Official Journal of the European Union}, 2024.

\bibitem{gdpr2016} European Parliament and Council, ``Regulation (EU) 2016/679
(General Data Protection Regulation),'' \emph{Official Journal of the European
Union}, 2016.

\end{thebibliography}
\end{document}